# Electrically tunable, rapid spin-orbit torque induced modulation of colossal magnetoresistance in $Mn_3Si_2Te_6$ nanoflakes

*Cheng Tan[1,2,‡], Mingxun Deng[3,‡], Yuanjun Yang[1,‡], Linlin An[1], Weifeng Ge[1], Sultan Albarakati[2,4], Majid Panahandeh-Fard[2], James Partridge[5], Dimitrie Culcer[6], Bin Lei[7], Tao Wu[8,9], Xiangde Zhu[10,*], Mingliang Tian[7,10], Xianhui Chen[8,9], Rui-Qiang Wang[3,*] and Lan Wang[1,5,*]*

[1] Lab of Low Dimensional Magnetism and Spintronic Devices, School of Physics, Hefei University of Technology, Hefei, Anhui 230009, China.

[2] ARC Centre of Excellence in Future Low-Energy Electronics Technologies (FLEET), School of Science, RMIT University, Melbourne, Victoria 3001, Australia.

[3] Guangdong Provincial Key Laboratory of Quantum Engineering and Quantum Materials, SPTE, South China Normal University, Guangzhou, Guangdong 510006, China.

[4] Physics Department, Faculty of Science and Arts, University of Jeddah, P.O. Box 80200, 21589Khulais, Saudi Arabia.

[5] Physics, School of Science, RMIT University, Melbourne, Victoria 3001, Australia.

[6] School of Physics and ARC Centre of Excellence in Future Low-Energy Electronics Technologies, UNSW Node, University of New South Wales, Sydney, New South Wales 2052, Australia.

[7] School of Physics and Optoelectronic Engineering, Anhui University, Hefei, Anhui 230601, China.

[8] Hefei National Laboratory for Physical Sciences at the Microscale, University of Science and Technology of China, Hefei, Anhui 230026, China.

[9] CAS Key Laboratory of Strongly-coupled Quantum Matter Physics, Department of Physics, University of Science and Technology of China, Hefei, Anhui 230026, China.

[10] Anhui Province Key Laboratory of Condensed Matter Physics at Extreme Conditions, High Magnetic Field Laboratory, HFIPS, Chinese Academy of Sciences (CAS), Hefei, Anhui 230031, China.

[‡] These authors equally contributed to the paper.

[*] Corresponding authors. Correspondence and requests for materials should be addressed to X. Z. (email: xdzhu@hmfl.ac.cn) R. W. (email: wangruiqiang@m.scnu.edu.cn) and L. W. (email: wanglan@hfut.edu.cn).

ABSTRACT: As a quasi-layered ferrimagnetic material, $Mn_3Si_2Te_6$ nanoflakes exhibit magnetoresistance behaviour that is fundamentally different from their bulk crystal counterparts. They offer three key properties crucial for spintronics. Firstly, at least $10^6$ times faster response comparing to that exhibited by bulk crystals has been observed in current-controlled resistance and magnetoresistance. Secondly, ultra-low current density is required for resistance modulation (~ 5 $A/cm^2$). Thirdly, electrically gate-tunable magnetoresistance has been realized. Theoretical calculations reveal that the unique magnetoresistance behaviour in the $Mn_3Si_2Te_6$ nanoflakes arises from a magnetic field induced band gap shift across the Fermi level. The rapid current induced resistance variation is attributed to spin-orbit torque, an intrinsically ultra-fast process (~nanoseconds). This study suggests promising avenues for spintronic applications. In addition, it highlights $Mn_3Si_2Te_6$ nanoflakes as a suitable platform for investigating the intriguing physics underlying chiral orbital moments, magnetic field induced band variation and spin torque.

Electrically controlled magnetism[1-4] has been a long-term pursuit in the field of spintronics.[5,6] With electrically controlled magnetic states, various spintronic devices, such as spin-orbit torque devices[7,8] and spin field effect transistors[9,10] can be realized without externally applied magnetic fields. This is vital in realizing ultra-high density magnetic memory and logic devices. Recently, a colossal magnetoresistance (MR) material $Mn_3Si_2Te_6$ has emerged and attracted significant attention.[11-20] $Mn_3Si_2Te_6$ is a quasi-layered ferrimagnetic material in which a perpendicularly applied magnetic field (along $z$-axis) can reduce the $xy$-plane resistivity by up to 7 orders of magnitude. An additional extraordinary characteristic of $Mn_3Si_2Te_6$ is that its magnetization and resistance can be tuned over a wide range by current flow.[18] However, this current-manipulated magnetization and resistance is a sufficiently slow process in bulk single crystalline $Mn_3Si_2Te_6$ to limit its application.[18] To date, only bulk single crystalline $Mn_3Si_2Te_6$ has been investigated experimentally. Research on nanoflake or thin film forms of $Mn_3Si_2Te_6$ is lacking. The mechanisms underlying colossal MR and current controlled magnetization and resistance are not fully explored, although the topological nodal-line degeneracy of spin- polarized bands[15] and a new quantum state known as chiral orbital currents[18] have been proposed as sources of these exotic properties. Here, we report that the colossal MR of $Mn_3Si_2Te_6$ nanoflakes is fundamentally different from the counterpart of bulk crystals. Furthermore, the resistance can be tuned by current over orders of magnitude and at least $10^6$ times more rapidly than that of bulk. These characteristics are thus in stark contrast to those of bulk single crystals. Finally, we have observed that the colossal MR can also be tuned via gate voltage. Colossal MR that is tunable via both current and applied gate voltage has never been observed and has exciting implications for next generation spintronic devices.

The $Mn_3Si_2Te_6$ lattice structure is illustrated in Figure 1a. The optical image in the lower right of Figure 1a shows a $Mn_3Si_2Te_6$ nanoflake (device #1, 140-nm thick). Based on prior research work and our measurement results (see Figure S2), bulk $Mn_3Si_2Te_6$ exhibits a monotonically decreasing colossal MR with increasing magnetic field. As shown in Figure 1b, the MR characteristics of the nanoflake device #1 significantly differ from those of $Mn_3Si_2Te_6$ crystals. At lower temperatures, unlike the monotonically decreasing MR observed in bulk $Mn_3Si_2Te_6$ (when applied magnetic field B//z-axis) with a single peak at zero field, two peaks appear symmetrically at positive and negative field. At 3 K, a flat valley exists in the MR from B=-4.5 T to 4.5 T. Further increasing the magnitude of the magnetic field results in MR peaks at ± 6 T that are one order of magnitude higher than the resistance in the valley. The resistance consistently decreases with higher magnetic fields up to ± 9 T. It should be emphasized that the valley-to-peak resistance ratios at low temperatures vary between devices from 1 (no peak) to $10^2$ (see Figure S8). Here, we define the applied magnetic field corresponding to the MR peak as $B_{peak}$. As shown in Figure 1c, $B_{peak}$ decreases with increasing temperature when T < $T_C$ (~76 K). The resistance at $B_{peak}$ ($R_{xx}^{max}$) decreases with increasing temperature when T < 20 K, while the width of the MR peak becomes larger with increasing temperature, indicative of a thermal agitation-related effect. As the temperature is raised from 20 K to 50 K, $R_{xx}^{max}$ grows and two distinct peaks merge into a single broad peak between 40 K and 50 K. Beyond 50 K, $R_{xx}^{max}$ decreases as the temperature is increased.

Temperature-dependent $R_{xx}$ curves (Figure 1d) also exhibit interesting behaviours. Above $T_C$, the resistance of the $Mn_3Si_2Te_6$ nanoflake increases sharply with decreasing temperature, as in a semiconductor. Below $T_C$, the peaks of $R_{xx}(T)$ curves shift to lower temperatures with increasing applied magnetic fields. This temperature-dependent resistance behaviour below $T_C$ is a temperature-scanned representation of the resistance profile in Figure 1b. As shown in Figure 1c, $B_{peak}$ increases monotonically from 0 T to 6.1 T as temperature decreases from $T_C$ to 3 K. Thus, while measuring the $R_{xx}$ (T) curve under a fixed magnetic field, the applied field matches the $B_{peak}$ (T) at a specific temperature, leading to a peak in resistance at this temperature.

The angular dependent MR curves of device #1 at B = 4- 9 T, I = 10 nA and T = 3 K are shown in Figure 2a. The measurement involved applying a fixed magnetic field then rotating the devices in a full 360 degrees, as depicted in the inset of Figure 2a. When $\theta$ = 0° (360°) and 180°, the magnetic field is parallel to the z-axis. When $\theta$ = 90° and 270°, the magnetic field is in the xy-plane and parallel to the applied current direction. The angular MR of device #1 is near-flat when $\theta$ is within 50°- 130° and 230°- 310° (around ± 40° away from the xy-plane). Interestingly, angular MR is evident around 180° and 0° (360°). In the higher field regime (6.5 T - 9 T), the angular MR shows a valley at 180º and 0° (360°) and two peaks near 180° and 0° (360°). In the lower field regime (B ≤ 6 T), the angular MR only shows a peak at 180º and 0° (360º). This experimental result is strongly related to the peaks and valleys of $R_{xx}$ (B) curve in Figure 1b. In the large magnetic field regime ($\geqslant$ 6.5 T), the magnetic field projected to the z-axis is large enough to induce the peak near $\theta$ = 0° (360°) and 180° (Figure 2a) and the valley at $\theta$ = 0° (360°) and 180°. In the small magnetic field regime ($\leqslant$ 6 T), the magnetic field projected to the z-axis is always smaller than the $B_{peak}$ in Figure 1b and reaches the highest value at $\theta$ = 0° (360°) and 180º, peaks therefore exist only at these angles. The same principle can be used to explain the angular-dependent MR at various temperatures with B = 9 T, as shown in Figure 2b. The peak and valley of the angular-dependent MR is strongly related to the peaks and valleys of the $R_{xx}$ (B) curves shown in Figure 1b, which demonstrates that the magnetic field projection to z-axis plays a key role in the angular magnetoresistance characteristics.

To further understand the relationship between the MR and the orientation of the applied magnetic field, we measured MR at fixed $\theta$ values. As shown in Figure 2c, $B_{peak}$ increases when the z-axis is tilted away from the orientation of the applied magnetic field (increasing $\theta$ value). In Figure 2d, the relationship between the MR and the field projected to z-axis (Bcos$\theta$) is plotted. If the MR originated solely from the magnetic field along the z-axis, we would expect to see a universal curve in the $R_{xx}$ (Bcos$\theta$) plots. However, as shown in Figure 2d, the non-universal curves clearly demonstrate that the MR of the nanoflakes originates from not only the magnetic field projection along the z-axis but also its projection in the xy-plane. This is a surprising result considering that the MR of bulk $Mn_3Si_2Te_6$ crystal under an in-plane magnetic field is eight orders of magnitude smaller than that observed when the same magnetic field is applied along the z-axis.[15] Additionally, the in-plane MR of device #1 is notably small (see Figure S3).

As previous research has revealed, bulk $Mn_3Si_2Te_6$ shows current-dependent MR, which is believed to originate from chiral orbital current.[18] As in bulk $Mn_3Si_2Te_6$ crystals, the MR of $Mn_3Si_2Te_6$ nanoflake can be strongly manipulated by the density of applied current as well. Figure 3a displays the current-dependent MR curves for device #1 in the positive field regime. Note that due to the compliance limit of the measurement system, the maximum applied current on device #1 can only reach around 50 nA. With an increase in the applied current, the resistance valley between the two peaks increases by one order of magnitude. As a result, the dual-peak MR vanishes at I = 50 nA and is replaced by a high resistance plateau between ±$B_{peak}$. The current-dependent MR evolution is clearer in another device (device#2) with two higher MR peaks (see Figure S5), where the maximum current cannot bring the resistance valley between the two peaks to a plateau. When the magnetic field is higher than $B_{peak}$, the current almost has no effect on resistance. The experimental results indicate that an applied current can generate a similar effect to that of a magnetic field. To better understand the current-dependent MR, we applied currents of 10, 30 and 50 nA with a 50-μs measurement delay to explore the swiftness of the device #1 $R_{xx}$ response at 0 T, as illustrated in Figure 3b. In these measurements, $R_{xx}$ followed the current with a response time of less than 50 μs (the limit imposed by our electronic meters). This current induced modulation of resistance in the $Mn_3Si_2Te_6$ nanoflakes is at least $10^6$ times faster than that recorded in bulk single crystals, which typically show a resistance relaxation time of tens of seconds.[18]

To further investigate the effect of the Fermi level shift on MR, we applied the protonic gating on $Mn_3Si_2Te_6$ nanoflakes, as shown in Figure 3c. A platinum contact serves as the bottom gate electrode and enables application of gate voltages adjacent to the solid proton conductor (SPC) layer. The generated electric field then drives the protons into the $Mn_3Si_2Te_6$ nanoflake. Generally, protonic gating enables substantial charge tunability, over two orders of magnitude higher (up to $10^{22}$ cm$^{-3}$)[21,22] than normal dielectric layers. As shown in Figure 3d, the MR of device #5 with a magnetic field applied along the *z*-axis at 2 K is significantly modulated by the applied gate voltage ($V_g$). At $V_g$ = 0 V, device #5 exhibited a flat resistance plateau, resembling the large current induced MR plateau in Figure 3a. Applying $V_g$ = -4 V resulted in a standard dual-peak MR with $B_{peak}$ = 5.35 T. Between $V_g$ = -4 and -8 V, the MR curves remain almost unchanged but a pronounced change occurs as $V_g$ exceeds -9 V. Further increasing the gate voltage, the $R_{xx}^{max}$ decreases and $B_{peak}$ shifts to lower values till 4.5 T at $V_g$= -12 V.

To further explain the experimental results, a model was constructed. As shown in Figure 1a, $Mn_3Si_2Te_6$ adopts a self-intercalated, quasi-layered structure with trigonal symmetry. Based on symmetry analysis, the effective Hamiltonian for the Te layer can be obtained (see Equation A1 in Supporting Information). From the effective Hamiltonian with the current-induced spin-orbit torque included, the orbital magnetic moment via coupling with the magnetic field and the carrier spins was calculated. The results demonstrate that the orbital magnetic moment couples only with the $z$ component of the magnetic field, and the MR is thus strongly dependent on the orientation of any applied magnetic field.

Figure 1 to Figure 3 display four key experimental results, namely, the large dual-peak of MR (Figure 1b), the angular dependent evolution of $R_{xx}$ ($B\cos\theta$) (Figure 2d), the current induced modulation of MR (Figures. 3a-b), and the effect of Fermi level shift on MR (Figure 3d). Using our model and including the effect of temperature on domain size, the experimental results can be explained as follows. The calculated evolution of the dispersion with the applied magnetic field is shown in Figure 4). It shows that with increasing $z$ component of the magnetic field, the band gap will move below the Fermi level, and thus produce a peak structure in the MR curves. As the magnetic field changes sign, the energy bands will interchange for the two chiralities. Two MR peaks, symmetrically distributed with respect to B (B//$z$-axis), are expected as a result. This is consistent with the low temperature measurements (3 K - 25 K) in Figure 1b. To further demonstrate this point, we calculate the longitudinal conductivity by the Kubo formula (see Equation A23 in Supporting Information). The numerical results for the resistivity are plotted as Figs. 4b-e. The calculated curves in Figure 4e only show the peaks as labelled by arrows in Figure 1d and do not include the ferrimagnetism-paramagnetism phase transition induced resistance peaks occurring at 75 K - 100 K. Also, the model does not consider any thermal broadening effects. Hence, Figure 4b agrees with the experimental results only at lower temperatures (Figure 1b, 3 K-25 K). In this temperature regime, both the $B_{peak}$ and the $R_{xx}^{max}$ decrease with increasing temperature. The higher temperature (> 30 K) MR can be explained by the evolved multidomain structure in nanoflakes (discussed in Supporting Information 1.4).

Because of the intrinsic molecular field, the magnetization direction of local spins is not collinear with external magnetic field B, so that the in-plane components of B can also affect the magnetization direction. Hence, the $R_{xx}$ (Bcosθ) curves do not form a universal line as shown in Figure 2d. The results of our calculations shown in Figure 4d are consistent with the experimental results. As mentioned above, due to the spin-orbit interaction, the current can induce spin polarization of the itinerant electrons. These will then exert a torque on the local spins and alter the direction of the magnetization of the Mn atoms. Therefore, the MR depends both on the azimuth of the magnetic field and the magnitude of the applied current. The calculated electric field (current) dependent magnetoresistance curve is shown in Figure 4c, which corresponds well with the experimental results (Figure 3a). Our current modulation measurements demonstrate that the current-induced resistance variation time is less than 50 μs, limited by the capability of our instruments. This is at least 6 orders of magnitude faster than the flip in magnetization and resistance relaxation reported in bulk $Mn_3Si_2Te_6$ with a multi-domain structure.[18] The dimension of device #1 is 9.7 μm × 6.5 μm × 140 nm. Hence, as indicated in Figure 3a, a current density of 5.5 A/cm$^2$ generates an effect of ~5 T magnetic field in device #1 and can generate a change of magnetic states and very large resistance variation. This current density is at least four orders of magnitude smaller than the switching current density required in conventional spin-orbit torque devices. Thus, ultra-low current density can induce rapid switching of magnetization, very large resistance variation and significantly modulated colossal magnetoresistance in nanoflake $Mn_3Si_2Te_6$.

The gate-dependent MR results can be interpreted as well. Based on the model, the initial position of the Fermi level at B = 0 affects the MR. If the Fermi level is initially in the gap, the MR curve should include a high-resistance plateau without two peaks, which corresponds to $V_g=$ 0 V in Figure 3d (device #5) and Figure S8b. If the Fermi level is in the conduction band or valence band, the MR will exhibit symmetric dual peaks at a positive field or a negative field, respectively. The ratio of the resistance at $B_{peak}$ and the resistance in the valley is determined by the density of states at the Fermi level (B=0). In Figure 3d, when a negative voltage is applied to the gate of device #5, negatively charged hydrogen ions are driven into the nanoflake, pushing the Fermi level from the band gap into the conduction band and resulting in the evolution of MR.

In conclusion, $Mn_3Si_2Te_6$ nanoflakes have produced colossal magnetoresistance that differ significantly from those of bulk single crystals. The current induced modulation of resistance observed in the $Mn_3Si_2Te_6$ nanoflakes is at least 6 orders of magnitude faster than that reported from the bulk single crystals. A current density of only 5.5 A/cm$^2$ produced significant modulation of the resistance of the nanoflakes. This current density is at least $10^4$ times smaller than the threshold current density in normal spin torque devices. The colossal magnetoresistance of $Mn_3Si_2Te_6$ nanoflakes can also be effectively manipulated by application of a gate voltage. Our theoretical calculations indicate that the current induced variation of MR and magnetization is due to a spin torque effect, which is an inherently ultra-fast (nanosecond scale) process. The unique combination of colossal MR, electrical gating as a means to tune MR and its ultra-fast and ultra-low current density tunability make this material extremely promising for spintronic applications. Further investigation of the chiral orbital moments, magnetic field induced band variation, and spin torque may reveal additional, exploitable phenomena.

**Experimental Section**

Crystal growth

$Mn_3Si_2Te_6$ single crystals were grown by slowly cooling from the melt. Mn pieces (4N), Si sponges(5N), and Te (6N) shots/crystals in mole ratio of 3:2:6 were loaded and sealed under vacuum in a quartz tube with a conical tip. The quartz tube was vertically placed in a box muffle furnace with the conical tip oriented to the bottom. The tube was heated to 1373 K where it remained for 6 hours. It was then cooled to 1123 K at a rate of 1.5 K/hour before final cooling.

Device fabrication

$Mn_3Si_2Te_6$ nanoflakes were mechanically exfoliated onto $SiO_2$/Si substrates in a glove box where oxygen and water levels were below 0.1 parts per million. Atomically smooth flakes were identified by an optical microscope in the glove box. Then the nanoflakes were transferred via a polycarbonate-based pick-up method onto pre-patterned electrodes or proton conductors for subsequent electron beam lithography and metal deposition processes. When a device was removed from the glove box, it was covered with a PMMA layer or PMMA droplet to prevent exposure to moisture and oxygen. To prepare solid protonic electrolyte, we first mixed tetraethyl orthosilicate (from AlfaAesar), ethanol, deionized water, phosphoric acid (as a proton source, from AlfaAesar, 85% wt%) with a typical molar ratio 1:18:6:0.03, then the mixed solution was stirred for 2 hours and annealed for another 2 hours at 50 °C in a sealed bottle to form polymerized Si-O-Si chains. Finally, the substrates (on which bottom gate electrodes were fabricated) were spin-coated with the prepared protonic solution and cured by baking at 150 °C for 25 mins. This yielded the solid proton conductor layer.

Electrical measurements

The electrical measurements were performed in a PPMS EverCool II system (Quantum Design, San Diego, CA, USA) with a base temperature of 1.8 K and a magnetic field of up to 9 T. The AC current was applied by a Keithley 6221 current source and the $V_{xx}$ signals were measured by SR560 low noise voltage preamplifier and SR830 lock-in amplifier (Stanford Research Systems). The gate voltages were applied by a Keithley 2450 Source Measure Unit (SMU). The SMU was also utilized for dc current switch measurements with a measurement delay time of 50 μs.

**Theoretical Calculations**

To model the magneto-transport of $Mn_3Si_2Te_6$, theoretically, we obtain an effective low-energy Hamiltonian for the Te layers based on the symmetry analysis. The quantum properties of the local Mn spins, namely, quantum correlation and thermal fluctuation effects, are taken into account. The spin-spin correlation between the local Mn atoms is captured by the Heisenberg model, where the spin density operators are obtained from the Holstein-Primakoff transform. Due to the thermal fluctuation, the magnitude of the spin density can be temperature dependent. The current-induced spin-orbit torque is derived from the Bloch equation. By diagonalizing the model Hamiltonian, we can arrive at the electron dispersion and wavefunction, by which we can further evaluate the orbital magnetic moment. The external magnetic field, via coupling with the orbital magnetic moment, can shift the energy band integrally, making the resistivity depend strongly on the magnetic field. The resistivity is calculated by means of the Green's function, which enters the Kubo-Streda formula, during which the magnon and impurity scattering effects can be included. Further details may be found in the Supporting Information.

FIGURES

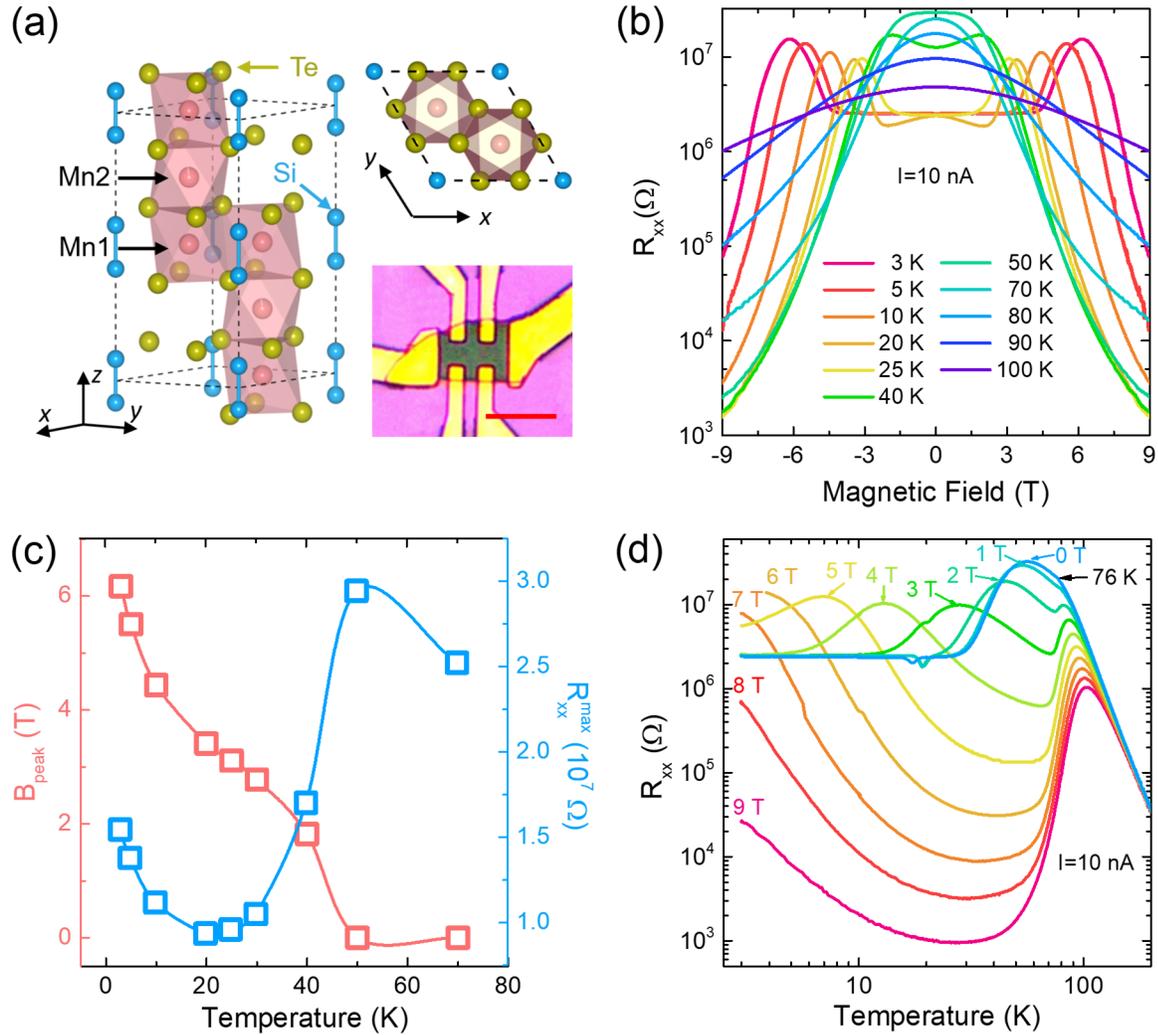

**Figure 1.** Electronic transport of a Mn$_3$Si$_2$Te$_6$ nanoflake device (device #1). (a) Crystal structure (upper left and right) and device image (lower right) of Mn$_3$Si$_2$Te$_6$ nanoflake. The red scale bar represents 10 μm. (b) MR at various temperatures at T= 3 K. (c) Temperature dependent B$_{peak}$ and $R_{xx}^{max}$ below T$_C$. (d) Temperature dependent R$_{xx}$ curves from B= 0 T to 9 T.

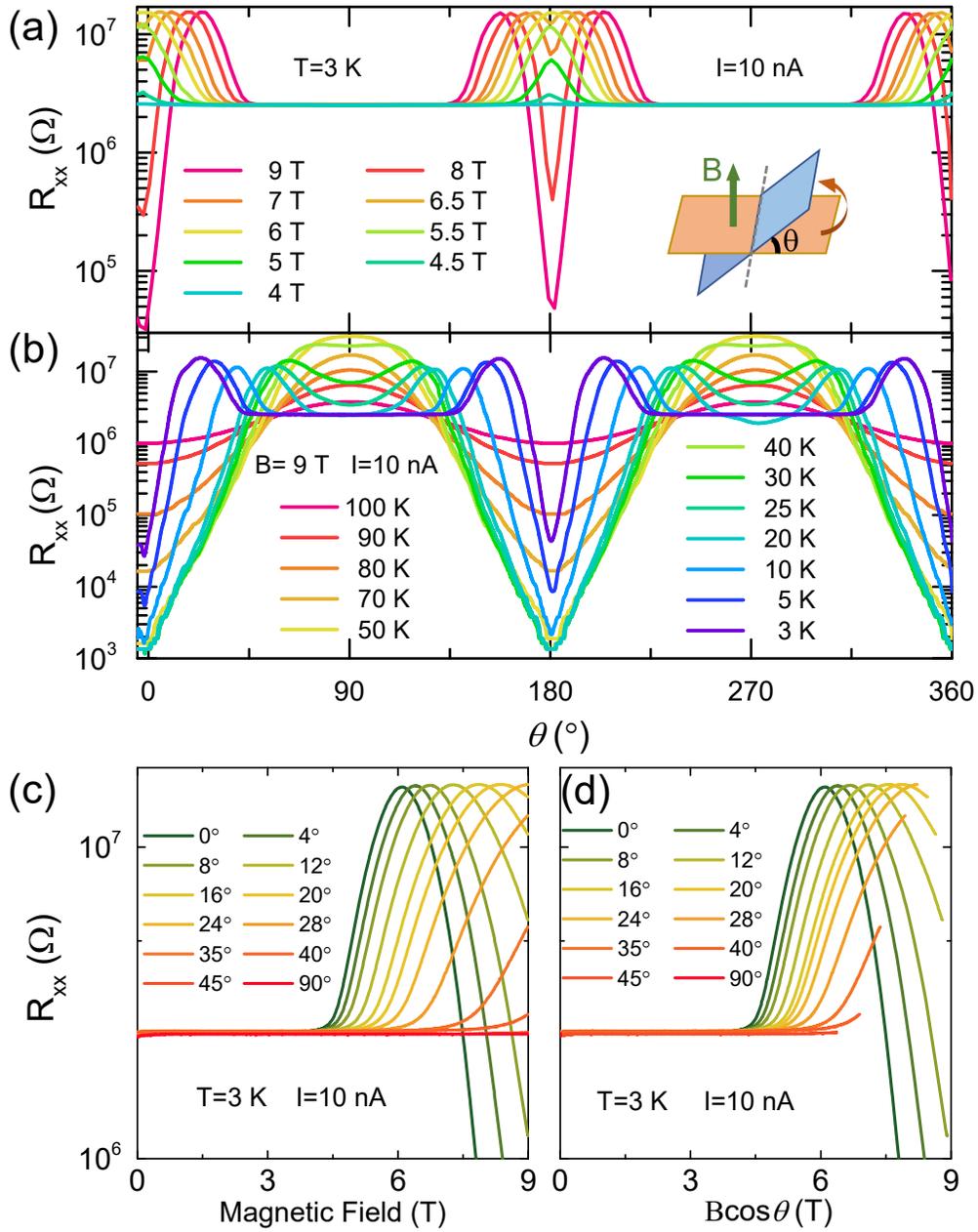

**Figure 2.** Angle dependent measurements on device #1. (a) $R_{xx}(\theta)$ curves at T= 3 K under various magnetic fields. (b) Temperature dependent $R_{xx}(\theta)$ curves at B = 9 T. (c) MR curves at fixed $\theta$ at T= 3 K. (d) $R_{xx}(B\cos\theta)$ curves derived from (c).

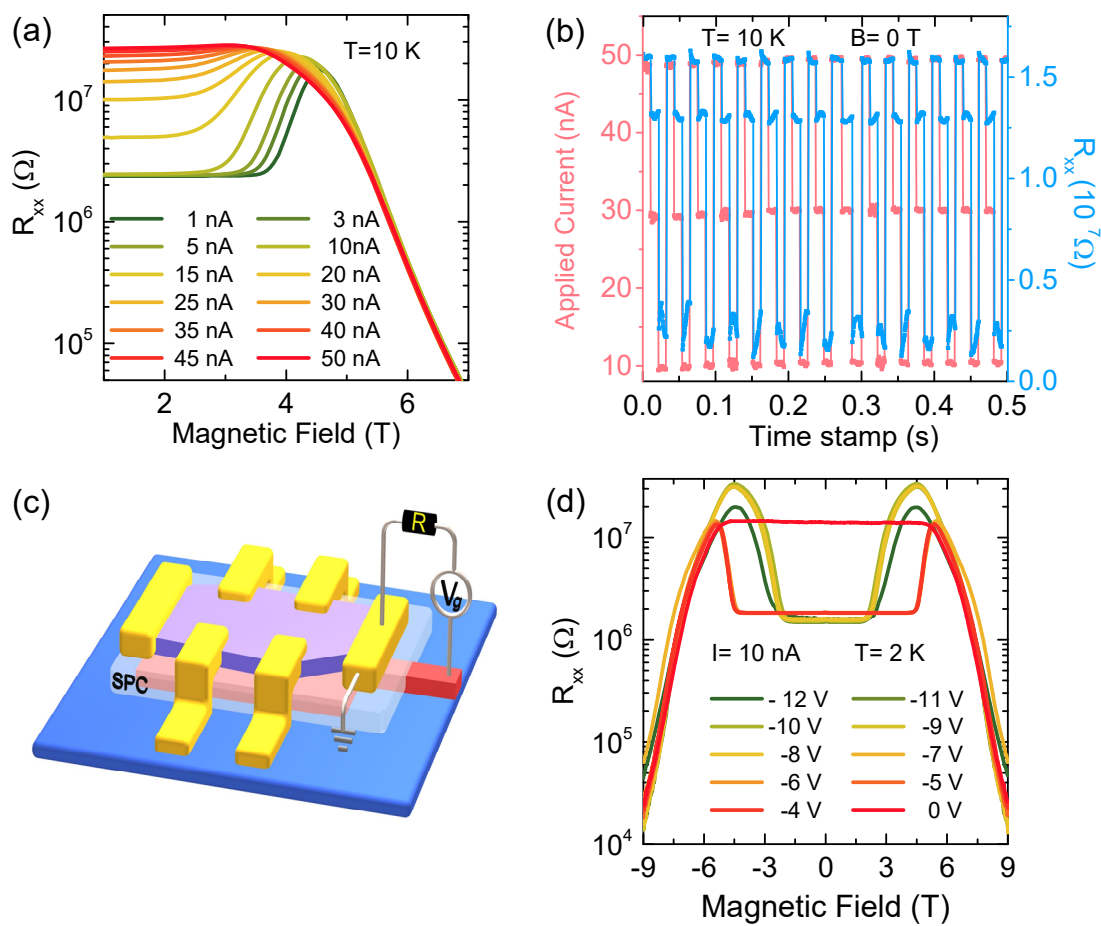

**Figure 3.** Current dependent (device #1) and gate voltage controlled (device #5) MR. (a) Current-dependent MR curves at T=10 K. (b) Switchable current modulated $R_{xx}$ at B = 0 T. The sequentially applied currents in one cycle are 50 nA, 30 nA and 10 nA. (c) Measurement configuration of a protonic gating device. (d) Gate voltage dependent MR curves at T= 2 K.

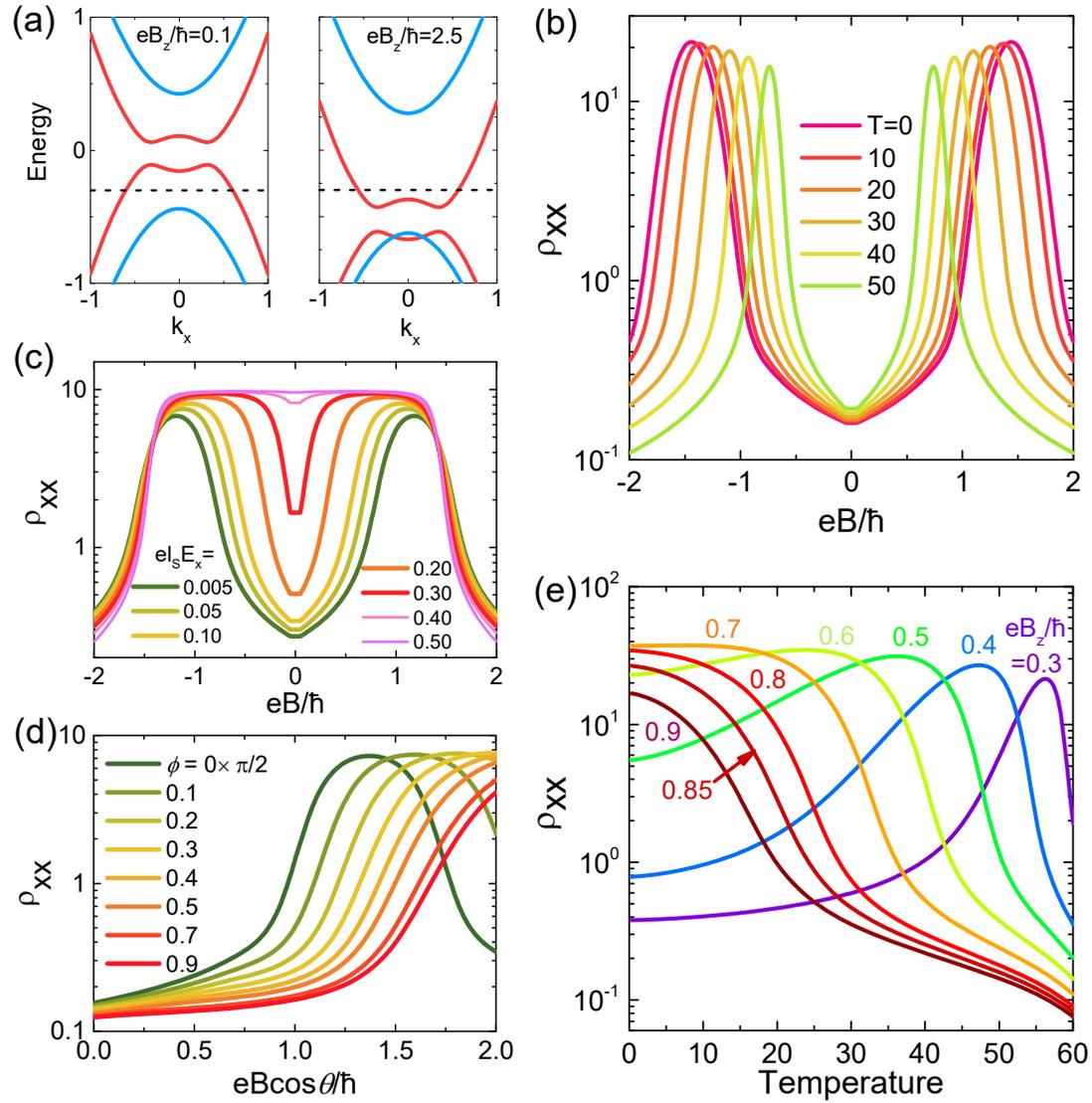

**Figure 4.** Theoretical model. Calculations of (a) dispersion with red (blue) corresponding the positive (negative) chirality, (b) $\rho_{xx}$ as a function of magnetic field at various temperatures, (c) $\rho_{xx}$ vs. magnetic field for varied current, (d) $\rho_{xx}$ vs. the $z$ component of the magnetic field for different azimuth angles ($\phi$) and (e) $\rho_{xx}$ against temperature at various magnetic fields. The parameters are set as $\hbar v_F = 0.25$, $m_0 = 0.15$, $m_1 = 1$, $c_0 = 0$, $c_1 = 0$, and $JS = 0.3$, respectively.

ASSOCIATED CONTENT

The Supporting Information is available free of charge.

Experimental section; magnetic measurements for single crystal; electronic transport measurements for single crystal; additional electronic transport measurements for device #1; multi-domain structure induced MR behaviour in device #1 when T > 30 K; transport measurements for other devices; additional measurements for proton-gated devices; Hall effect measurements; theoretical calculation; model Hamiltonian; current-induced spin-orbit torque; orbital magnetic moment; magneto-transport

AUTHOR INFORMATION

**Corresponding Authors**

*Xiangde Zhu (email: xdzhu@hmfl.ac.cn)

*Rui-Qiang Wang (email: wangruiqiang@m.scnu.edu.cn)

*Lan Wang (email: wanglan@hfut.edu.cn)

**Author Contributions**

R.-Q.W. and L.W. conceived the project. L.A. and X.Z. synthesized the single crystals. C.T. fabricated the devices and performed the experimental measurements, assisted by Y.Y., W.G., B.L., S.A., M.P.-F. and T.W.. M.D. and R.-Q.W. provided theoretical calculations. C.T., M.D., J.P., Y.Y., D.C., X.H.C., M.T., R.-Q.W. and L.W. analyzed the data and wrote the manuscript with assistance from all authors. ‡These authors contributed equally.

**Funding Sources**

This research was supported by the National Natural Science Foundation of China (12374177 and 52072102), the Funding for Infrastructure and Facility of Hefei University of Technology and the Australian Research Council Centre of Excellence in Future Low-Energy Electronics Technologies (Project No. CE170100039).

**Notes**

The authors declare no competing financial interest.

ACKNOWLEDGMENT

This research was performed in part at the RMIT Micro Nano Research Facility (MNRF) in the Victorian Node of the Australian National Fabrication Facility (ANFF) and the RMIT Microscopy and Microanalysis Facility (RMMF).

ABBREVIATIONS

MR, magnetoresistance; SPC, solid proton conductor.